\documentclass[a4paper,12pt]{article}

\usepackage{a4,amsfonts,amsmath,amssymb}
\usepackage{cite}
\usepackage{url}
\usepackage{hyperref}
\def\be{\begin{equation}}
\def\ee{\end{equation}}
\def\nn{\nonumber}

\def\D{\mathcal{D}}
\def\H{\mathcal{H}}
\def\L{\mathcal{L}}
\def\M{\mathcal{M}}
\def\N{\mathcal{N}}
\def\Q{\mathcal{Q}}
\def\SS{\mathcal{S}}
\def\V{\mathcal{V}}

\def\at{\tilde{a}}
\def\bt{\tilde{b}}
\def\et{\tilde{e}}
\def\tt{\tilde{t}}
\def\It{\tilde{I}}
\def\Jt{\tilde{J}}
\def\Kt{\tilde{K}}
\def\rt{\tilde{r}}
\def\xt{\tilde{x}}
\def\yt{\tilde{y}}
\def\zt{\tilde{z}}

\def\d{\partial}
\title{
{\baselineskip -.2in
\vbox{\small\hskip 4in \hbox{IITM/PH/TH/2012/8}}
\vbox{\small\hskip 4in \hbox{IMSc/2012/6/13 \quad \quad}}
} 
\vskip .4in
Generalized Attractors in Five-Dimensional  Gauged Supergravity \\}
\date{}
\author{Karthik Inbasekar$^a$ and Prasanta K. Tripathy$^b$\\ \\
	\it $^a$Institute of Mathematical Sciences, \\
	\it CIT Campus, Taramani,\\
	\it Chennai 600113, India.\\
	\\ \\
	\it $^b$Department of Physics,\\
	\it Indian Institute of Technology Madras,\\
	\it Chennai 600 036, India.\\ \\
	email: $^a$ikarthik@imsc.res.in ; $^b$prasanta@physics.iitm.ac.in \\ \\}

\begin{document}

\maketitle
\begin{abstract}

In this paper we study generalized attractors in $\N = 2$ gauged supergravity theory in five dimensions 
coupled to arbitrary number of hyper, vector and tensor multiplets. We look for attractor solutions with constant
anholonomy coefficients. By analyzing the equations of motion we derive the attractor potential. We further show 
that the generalized attractor potential can be obtained from the fermionic shifts. We study some simple examples 
and show that constant anholonomy gives rise to homogeneous black branes in five dimensions. 

\end{abstract}

\newpage
\section{Introduction}

The attractor mechanism plays a crucial role in understanding the origin of black hole entropy  in supergravity theories.
Originally the attractor mechanism was realized in the context of four dimensional $N=2$ supergravity coupled to a 
number of vector multiplets\cite{Ferrara:1995ih}. For supersymmetry preserving black hole solutions the scalar 
fields in such theories run into a fixed point at the horizon. Subsequently it was realized that the attractor 
mechanism is a consequence of extremality  of the black hole \cite{Ferrara:1997tw} and that it can even occur in 
non-supersymmetric theories which need not have a supergravity origin \cite{Goldstein:2005hq}.

Recently there has been a renewed interest in the attractor mechanism because of its relevance in gravity duals to
field theories violating Lorentz symmetry. Attractor mechanism in gauged supergravity for static, supersymmetric black holes  was first studied in great detail in Ref.\cite{Cacciatori:2009iz}. Charged dilatonic black branes with Lifshitz-like near horizon geometry 
and asymptotic AdS, exhibiting attractor behavior have been analyzed \cite{Goldstein:2009cv,Goldstein:2010aw}. 
A large class of extremal homogeneous anisotropic  black branes have been extensively studied  and a 
classification of these solutions in terms of Bianchi Classes was given \cite{Iizuka:2012iv}.

These solutions possess an important common property, i.e., they have constant anholonomy coefficients. 
A general analysis of attractors with constant anholonomy coefficients in $\N=2$ gauged supergravity in four 
dimensions has been carried out by Kachru et.al. in \cite{Kachru:2011ps}. Such gauged supergravity theories 
are known to admit Lifshitz \cite{Cassani:2011sv} as well as Schrodinger  \cite{Halmagyi:2011xh} type solutions 
and in some special cases they can be embedded in string theory\cite{Balasubramanian:2010uk,Donos:2010tu,Gregory:2010gx,Donos:2010ax}. Ref. \cite{Kachru:2011ps} showed that under very general assumptions for the scalar as well as the vector fields these theories admit attractors. Near the attractor point the equations of motion become algebraic and the attractor potential takes a simple form. Moreover, due to constant anholonomy the components of the Riemann tensor become constant in tangent space. It follows that the curvature invariants are constant and hence the attractor geometries characterised by constant anholonomy coefficients are regular. 

Our goal in this paper is to generalise the analysis of \cite{Kachru:2011ps} to ${\N=2}$ gauged supergravity in 
five dimensions coupled to arbitrary number of hyper, vector and tensor multiplets. Such theories have richer 
structure because of the presence of tensor multiplets. Also, in most cases, the four dimensional $\N = 2$ 
gauged supergravity follows as a consistent truncation of the five dimensional counterpart. In ungauged
supergravity a large class of BPS solutions in four and five dimensions are related to each other, for 
example the multi-centered black holes in four dimensions can be lifted to a black ring in five dimensions
\cite{Gaiotto:2005xt}. Precise correspondences relating attractors in four and five dimensions will be left 
for future work. In this paper we merely generalise the analysis of \cite{Kachru:2011ps} and derive
the generalised attractor potential. Subsequently, using the formalism of generalised attractors we obtain 
some of the simplest black brane geometries such as the AdS Reissner-Nordstrom black hole as well as the Lifshitz metric.
In addition, in this paper we consider a simple gauged supergravity model in five dimensions with one 
vector and two tensor multiplets \cite{Gunaydin:2000xk}. We show that some of the Bianchi type solutions
considered recently in \cite{Iizuka:2012iv} can be realised as attractor solutions in this very simple supergravity model.

Before proceeding further, we would like to caution our readers about the usage of the term attractor in this 
context. The attractor mechanism originally studied in \cite{Ferrara:1995ih} was in the context of supersymmetry
preserving black hole configurations. As it has been subsequently realized, the critical points of the black hole 
effective potential may not be supersymmetric in general \cite{Ferrara:1997tw}.  A detail analysis of stability of  
non-supersymmetric black holes in asymptotic Minkowski space carried out in \cite{Goldstein:2009cv}
suggests that the stable attractors corresponds to the absolute minima of the effective black hole potential. 
This condition slightly differs for black holes in (Anti)deSitter spaces \cite{Goldstein:2009cv}. For the geometries
under consideration in the present work, we only analyze the critical points of the attractor potential leaving 
the stability analysis for future investigation \cite{karthik}.

The plan of the paper is as follows. In the following section we will discuss the most general $\N = 2$ gauged 
supergravity coupled to hyper, vector and tensor multiplets in five dimensions. Subsequently in $\S$\ref{generalisedattractors} we will
analyse the equations of motion for constant anholonomy coefficients and derive the attractor potential. In 
this section we also show that the attractor potential can independently be derived from the fermionic shifts.
In $\S$\ref{constantanholonomy} we consider some examples of attractors with constant anholonomy. Finally, we summarise our results
in $\S$\ref{summary}. We explain some of the notations, definitions and conventions in appendix $\S$\ref{appendix}. We summarise some details of the gauged supergravity model \cite{Gunaydin:2000xk} required for the later parts of section $\S$\ref{constantanholonomy} in appendix $\S$\ref{appendix1}. We give the gauged supergravity field equations for the Bianchi Type II and Type VI metrics in appendix $\S$\ref{bianchi26fieldeq}. 

\section{Gauged Supergravity}\label{gaugedsugra}

In this section we give a brief summary of the $\N = 2$ gauged supergravity theory in five dimensions. 
The five dimensional supergravity with a generic gauging of the symmetries of the scalar manifold and 
the $SU(2)_R$ R-symmetry was constructed by Ceresole and Dall'Agata \cite{Ceresole:2000jd}. In this 
section we will closely follow their notations and conventions. 
For some early work on gauged supergravity  see \cite{Gunaydin1985573,Gunaydin1984244,Gunaydin:2000xk,Gunaydin:1999zx} and for a review see \cite{Fre:2001jd}. 

The theory has the following field content  \cite{Ceresole:2000jd}:
\begin{itemize}
 \item The gravity multiplet 
 contains the graviton $e_\mu^a$, two gravitinos $\psi_\mu^i$ and the graviphoton $A_\mu$.
 \item The hypermultiplet 
 contains a doublet of fermions $\zeta^A$ with $A=1,2$ and four real scalars  $q^X$ with $X=1,\ldots,4$.
  \item The vector multiplet 
  contains a vector field $A_\mu$, $SU(2)_R$ doublet of fermions $\lambda^i$  and a real scalar field $\phi$. 
 \item The tensor multiplet 
 contains a massive antisymmetric self-dual tensor field $B_{\mu\nu}$, $SU(2)_R$ doublet of fermions $\lambda^i$ and a real scalar field $\phi$. 
\end{itemize}
To summarise, for $n_V$ vector, $n_T$ tensor and $n_H$ hypermultiplets the field content is given by
$\{ e_\mu^{\ a}, \psi_\mu^i, A_\mu^I, B_{\mu\nu}^M, \lambda^{i \at}, \zeta^{A} ,\phi^{\xt},q^X \}.$
The scalars in the vector and tensor multiplets are collectively denoted by $\phi^{\xt}$, where 
$\xt=1,2,\ldots ,n_v+n_T$. The vector field index is $I=0,1,\ldots, n_V$ and $I=0$ refers to the graviphoton. 
The index $M=1,2,\ldots, n_T$ counts the number tensor multiplets. The vector and tensor field strengths are 
collectively written as $\H_{\mu\nu}^{\It} = (F^I_{\mu\nu}, B^M_{\mu\nu})$ where $\It =(I,M)$. The gauginos $\lambda^{i \at}$ in the vector and tensor multiplets transform as vectors under $SO(n_V+n_T)$ and $\at= 1,2\ldots,n_V+n_T$ is a flat index. The quaternions $q^X, X=1,2,\ldots 4n_H$ are the scalars in the $n_H$ hypermultiplets. The hyperinos $\zeta^{A}, A=1,2,\ldots,2n_H$ form fundamental representations of $USp(2n_H)$ and $USp(2)\simeq SU(2)$. The conventions on the $SU(2)$ tensor $\epsilon^{ij}$ are summarised in appendix $\S$\ref{appendix}. 

\subsection{Symmetries of the scalar manifold}
The scalars in the theory parametrise a manifold that factorises into a direct product of a very special and quaternionic manifold \cite{deWit:1991nm}.
\be
\M_{scalar}=\SS(n_v+n_T) \otimes \Q(n_H) \ .
\ee
The manifold $\SS(n_v+n_T)$ is a very special manifold described by a $(n_v+n_T)$ dimensional hypersurface\cite{Gunaydin1984244}. The $n_V+n_T+1$ co-ordinates which parametrise this surface $h^{\It}=h^{\It}(\phi)$ satisfy the constraint,
\be\label{scalarcondition}
C_{\It \Jt \Kt} h^{\It} h^{\Jt} h^{\Kt}=1
\ee 
where $C_{\It \Jt \Kt}$ are constant symmetric tensors. This symmetric tensor also appears in the five-dimensional
Chern-Simons term. In order for the action to be supersymmetric and gauge invariant, the only non-vanishing 
components of the tensor $C_{\It\Jt\Kt}$  must be of the form $C_{IJK}$ and $C_{IMN}$ \cite{Gunaydin:1999zx}. 
The indices on the co-ordinates $h^{\It}$ on the hypersurface $\SS$ as defined by Eq.\eqref{scalarcondition} are
raised and lowered using the metric $a_{\It \Jt}$ \cite{Gunaydin1984244}:
\begin{align}\label{ambientmetric}
 a_{\It \Jt} &= h_{\It} h_{\Jt} + h^{\xt}_{\It}h^{\yt}_{\Jt} g_{\xt \yt} \nn \\
g_{\xt \yt} &= h^{\It}_{\xt} h^{\Jt}_{\yt} a_{\It \Jt}
\end{align}
where the metric $g_{\xt \yt}$ is defined through,
\begin{align}\label{scalarvielbein}
 f^{\at}_{\xt}f^{\bt}_{\yt}\eta_{\at \bt} &= g_{\xt \yt} \nn \\
 f^{\at}_{[\xt , \yt]}+\Omega^{\at \bt}_{[ \yt}f^{\bt}_{\xt ]} &=0 .
\end{align}
 Here $f^{\at}_{\xt}$ and $\Omega^{\at \bt}_{ \yt}$ are the $(n_V+n_T)$-bein and the spin connection on $\SS$ respectively.

The quaternionic K\"{a}hler manifold $\Q$ is parametrised by the scalars $q^X$. Since the holonomy group of $\Q$ is $USp(2) \otimes USp(2n_H)$ one introduces the vielbeins $f^X_{iA}$ and define the metric $g_{XY}$ on $\Q$ as follows (see, for example, Ref.\cite{Ceresole:2000jd}):
\begin{align}\label{quaternionvielbein}
f^X_{iC}f^{YC}_j+f^Y_{iC}f^{XC}_j &=g^{XY}\epsilon_{ij} \nn\\
g_{XY}f^X_{iA}f^Y_{jB} &=\epsilon_{ij}C_{AB} \nn \\
f^X_{iA}f^{Yi}_B + f^Y_{iA}f^{Xi}_B &=\frac{1}{n_H}g^{XY}C_{AB},
\end{align}
where $\epsilon_{ij}$ and $C_{AB}$ are $SU(2)$ and $USp(2n_H)$ invariant tensors respectively. 
\subsection{Gauging the symmetry group}
In this section we will review the gauging of a suitable subgroup $K$ of the isometry group $G$ of the full 
scalar manifold. The gauging can be described in terms of what is called as the momentum map
associated with the scalar manifold. For the $d=4, \N=2$ theories the scalar manifold is special K\"{a}hler and there exists a momentum map for the 
isometries (see section 7 of \cite{Andrianopoli:1996cm}). Whereas in the case of $d=5$ the scalar manifold is 
very special, real and non-symplectic. Hence there does not exist a momentum map for the isometries. 
However, the quaternion structure is the same in both $d=4$ and $d=5$ theories. Consequently there exist 
Killing prepotentials (i.e. there exist  Killing vectors which are given  in terms of the derivatives of prepotentials)
as in the 4d case. A significant difference in the 5d $\N=2$ theory is the presence of tensor multiplets which 
originate due to the gauging. The vectors in the ungauged theory, upon gauging fall into the following 
representations of $K$ in general\cite{Gunaydin1985573}:
\be
vector\rightarrow Adj(K)+ Singlets(K) + Non singlets(K).
\ee
For the singlets the structure constants of $K$ are assumed to be zero and if $K$ is abelian the presence of singlets do not change anything. The non-singlets are charged under $K$ and gauging them directly will lead to mass terms for the gauge fields which breaks supersymmetry. This issue is resolved by dualising the charged vectors to tensor fields \cite{Townsend198438}. Gauging $n_V+1$ vectors gives $n_T=dim(G)-n_V$ vectors charged under $K$, which are then dualised to tensor fields.  

Having identified the isometry group on $\M$, one introduces Killing vectors $K^{\xt}_I(\phi)$ and $K^X_I (q)$ that act on $\SS$ and $\Q$ ,
\begin{align}\label{gauging}
\phi^{\xt} & \rightarrow \phi^{\xt} +\epsilon^I K^{\xt}_I(\phi) \nn\\
q^X & \rightarrow q^X + \epsilon^I K^X_I (q),
\end{align}
where $\epsilon^I$ are infinitesimal parameters. Gauging the supergravity is done by replacing the ordinary 
derivatives on scalar and fermions by the $K$-covariant derivatives. The bosonic part of the 
theory then gets the following replacements \cite{Gunaydin:1999zx,Ceresole:2000jd}:
\begin{align}\label{bosoniccovariantderivative}
\d_\mu\phi^{\xt} & \rightarrow \D_\mu\phi^{\xt} \equiv \d_\mu\phi^{\xt} + g A^I_\mu K_I^{\xt}(\phi) \nn\\
\d_\mu q^X & \rightarrow \D_\mu q^X \equiv \d_\mu q^X + g A^I_\mu K_I^X (q) \nn \\
\nabla_\mu B^M_{\nu\rho} & \rightarrow \D_\mu B^M_{\nu\rho}\equiv \nabla_\mu B^M_{\nu\rho} + g A^I_\mu \Lambda^M_{IN} B^N_{\nu\rho},
\end{align}
where $g$ is the gauge coupling and $\nabla_\mu$ is the Lorentz covariant derivative. The $\Lambda^M_{IN}$ are constant matrices which are representations of $K$.

In addition to gauging the scalar manifold symmetries Ref.\cite{Ceresole:2000jd} also discussed the gauging of
 the $SU(2)_R$ symmetry. In this case, the $SU(2)_R$ connection $\omega^i_j$ is replaced by $\omega^i_j+ g_R A^I P_{I i}^{\ \ j}(q)$, where $g_R$ is the $SU(2)_R$ gauge coupling and $P_{I i}^{\ \ j}(q)$ are Killing prepotentials that exist due to the quaternionic structure on the hypermultiplet sector. In particular the Lorentz covariant 
 derivative acting on the gravitino is replaced by a $SU(2)_R$ gauge covariant derivative.
\be\label{gravitinocovariantderivative}
\nabla_\mu \psi_{\nu i} \rightarrow \nabla_\mu \psi_{\nu i} +g_R A_\mu^I P_{I i}^{\ \ j}(q) \psi_{\nu j}
\ee
The covariant derivatives on other fermions also get this $SU(2)_R$ connection piece in addition to the $K$ covariant pieces as in \eqref{bosoniccovariantderivative}, however we do not list them here. 

\subsection{Lagrangian}
The bosonic part of the five dimensional $\N=2$ gauged supergravity is given by \cite{Ceresole:2000jd}:
\begin{align}\label{lagrangian}
\hat{e}^{-1} \L^{\N=2}_{Bosonic}= & -\frac{1}{2}R -\frac{1}{4} a_{\It \Jt }\H_{\mu\nu}^{\It}\H^{\Jt \mu\nu}-\frac{1}{2}g_{XY} \D_\mu q^X \D^\mu q^Y
-\frac{1}{2}g_{\xt \yt}\D_\mu \phi^{\xt}\D^\mu\phi^{\yt}\nn \\
& +\frac{\hat{e}^{-1}}{6\sqrt{6}}C_{IJK}\epsilon^{\mu\nu\rho\sigma\tau}F^I_{\mu\nu}F^J_{\rho\sigma}A^K_\tau+\frac{\hat{e}^{-1}}{4g}\epsilon^{\mu\nu\rho\sigma\tau}\Omega_{MN}B^M_{\mu\nu}\D_\rho B^N_{\sigma\tau}\nn\\
&-\V(\phi,q).
\end{align}
Where $\hat{e}= \sqrt{-det g_{\mu\nu}}$ and $\Omega_{MN}$ is a constant real symplectic matrix that satisfies the following conditions
\be
\Omega_{MN}=-\Omega_{NM} ;\quad \Omega_{MN} \Omega^{NP}=\delta^P_M.
\ee
Gauging the supergravity introduces a non-trivial\footnote{In ungauged theories the scalars are just moduli and there is no such scalar potential.} scalar potential which is given by,
\be\label{potential}
\V(\phi,q)=2g^2 W^{\at}W^{\at}-g_R^2 [ 2P_{ij}P^{ij}-P^{\at}_{ij}P^{\at ij} ]+2g^2\N_{iA}\N^{iA}
\ee
where,
\begin{align}
 P_{ij} &\equiv h^I P_{Iij} \nn \\
 P^{\at}_{ij} &\equiv h^{\at I}P_{Iij} \nn \\ 
 W^{\at} &\equiv\frac{\sqrt{6}}{4}h^I K^{\xt}_I f^{\at}_{\xt} \nn\\
 \N^{iA} &\equiv\frac{\sqrt{6}}{4}h^I K^X_I f^{A i}_X. 
\end{align}
The bosonic part of the supersymmetry transformation rules are:
\begin{align}\label{susytransformation}
 \delta_\epsilon\psi_{\mu i} &=\sqrt{6} \nabla_\mu\epsilon_i +\frac{i}{4}h_{\It} (\gamma_{\mu\nu\rho}\epsilon_i - 4 g_{\mu\nu}\gamma_\rho\epsilon_i)\H^{\nu\rho\It}+ i g_R \gamma_\mu\epsilon^j P_{ij} \nn \\
\delta_\epsilon\lambda_i^{\at} &= -\frac{i}{2} f^{\at}_{\xt}\gamma^\mu\epsilon_i\D_\mu\phi^{\xt} +\frac{1}{4}h^{\at}_{\It}\gamma^{\mu\nu}\epsilon_i\H^{\It}_{\mu\nu}+ g_R \epsilon^j P^{\at}_{ij} +g W^{\at}\epsilon_i \nn\\
\delta_\epsilon\zeta^A &=-\frac{i}{2} f^A_{iX}\gamma^\mu \epsilon^i \D_\mu q^X +g \epsilon^i \N_i^A.
\end{align}
A supersymmetric ward identity relates the potential $\V(\phi,q)$, the gravitino mass matrix $P_{ij}$ and the fermionic shifts \cite{DeWit1982323,Ferrara1986600,Cecotti1985367,Fre:2001jd,Ceresole:2001wi}. As one can see from \eqref{susytransformation} the scalar potential \eqref{potential} can be written in terms of the squares of the gravitino mass matrix and the fermion shifts in the supersymmetry transformations that appear due to the gauging. Later, we show that the attractor potential can be written in a similar way.

\section{Generalised Attractors}\label{generalisedattractors}

In this section, we consider the $\N=2, d=5$ gauged supergravity coupled to vector, tensor and 
hypermultiplets and show that the equations of motion reduce to algebraic equations in the tangent space. 
We also show that at the attractor point the scalar equations reduce to an extermination condition of an attractor potential. Later, we construct the attractor potential from squares of the gravitino mass matrix and fermion shifts defined at the attractor point. This is a straightforward generalisation of the analysis done for $\N=2, d=4$ gauged supergravity by \cite{Kachru:2011ps} to the five dimensional case. For simplicity, we take the gauge group $K$ to be abelian, $K=U(1)^{n_V+1}$.

We consider attractors with the following ansatz for the scalar, vector and tensor fields at the attractor point:
\be\label{ansatz1}
\phi^{\zt}={\rm const} \ ; q^Z={\rm const} \ ; A^I_a={\rm const} \ ; B^M_{ab}={\rm const} \ ; c_{bc}^{\ \ a}={\rm const}
\ee
Here, in addition to the assumptions considered in \cite{Kachru:2011ps}, we take the tensor fields to be 
constant along the tangent space. As we will see, this is necessary in order to reduce the field equations to 
attractor equations (which are algebraic equations for all practical purposes). The constancy of $c_{bc}^{\ \ a}$ 
ensure the regularity of the resultant geometry and together with constant $A^I_a$, they ensure that the field 
strengths are constant (see Appendix \eqref{appendix}) at an attractor point, which is expected for an attractor 
behaviour. 

\subsection{Equations of Motion}\label{eom}
We now analyse the equations of motion of the bosonic fields in $\N=2, d=5$ gauged supergravity. We first derive the gauge field and tensor field equations. Subsequently we discuss the Einstein's equations followed by the equation of motion for the scalars
and quarternions which leads to the attractor potential.
\subsubsection{Gauge Fields and Tensor fields}
The Lagrangian for the $\N=2 ,d=5$ theory contains tensor fields and a Chern Simons term which contribute to the gauge field equation of motion:
\begin{align}\label{gaugefieldeq}
\d_\mu (\hat{e} a_{I \Jt}\H^{\Jt \mu \nu})= &-\frac{1}{2\sqrt{6}}C_{I \Jt \Kt}\epsilon^{\nu\mu\rho\sigma\tau} \H^{\Jt}_{\mu\rho}\H^{\Kt}_{\sigma\tau}\nn\\
&+ g \hat{e}\big[g_{XY} K^X_I \D^\nu q^Y + g_{\xt \yt} K^{\xt}_I \D^{\nu}\phi^{\yt}\big].
\end{align}
We have used the Bianchi identity $d(*F)=0$, the symmetry of $C_{\It \Jt \Kt}$\footnote{As mentioned earlier, the only non vanishing components are $C_{IJK}$ and $C_{IMN}$ , which can be used to write $C_{IJK}F^J F^K+C_{IMN}B^M B^N=C_{I \Jt \Kt}\H^{\Jt}\H^{\Kt}$.} and $\Omega_{MN}\Lambda^N_{IP}=\frac{2}{\sqrt{6}}C_{PMI}$ \cite{Gunaydin:1999zx} for simplification. 
Since the scalars, gauge fields, field strengths (see Appendix \eqref{appendix}) and tensor fields are constant at the attractor points \eqref{ansatz1} , the derivatives drop out 
\be
\d_a F^{abI}=0 ; \ \d_a A^{b I}=0 ; \ \d_a B^M_{bc}=0 ;\ \d_a \phi^{\zt}=0 ;\ \d_a q^Z=0.
\ee
In tangent space the gauge field equation can be written as,
\begin{align}\label{tangentgaugefieldeq}
\hat{e} \ a_{I \Jt}[\omega_{a,\ c}^{\ \ a} \H^{cb \Jt}+\omega_{a,\ c}^{\ \ b} \H^{ac \Jt}]=&-\frac{1}{2\sqrt{6}}C_{I \Jt \Kt}\epsilon^{b a c d e} \H^{\Jt}_{a c}\H^{\Kt}_{d e}\nn\\
&+ g^2 \hat{e} \big[g_{XY} K^X_I K^Y_J + g_{\xt \yt} K^{\xt}_I K^{\yt}_J \big] A^{J b}.
\end{align}
 The spin connection is expressed in terms of anholonomy coefficients in the absence of torsion \eqref{anholonomyandspinconnection}. Hence, constant anholonomy implies constant spin connection. Thus \eqref{tangentgaugefieldeq} is an algebraic equation at the attractor point.
 
 Similarly the tensor field equation can be worked out as,
\be\label{tensorfieldeq}
\frac{1}{g}\epsilon^{\mu\nu\rho\sigma\tau}\Omega_{MP}\D_\rho B^M_{\mu\nu}+\hat{e} a_{\It P} \H^{\It \sigma\tau}=0.
\ee
(Note that the Bianchi identity for the $B$-field, $d(B^M_{\mu\nu})\neq 0$ in general \cite{Gunaydin:1999zx}. 
So we will not be able to use it for simplification.) Converting the above to tangent space, we get
\begin{align}\label{tangenttensorfieldeq}
\frac{1}{g}\epsilon^{abcde}\big[c_{ac}^{\ \ f} B^M_{fb}+g A^I_c \Lambda^M_{IN}B^N_{ab} \big]\Omega_{MP} +\hat{e} a_{\It P} \H^{\It de}=0.
\end{align}
As in the previous case, the equation of motion \eqref{tangenttensorfieldeq} reduced to an algebraic equation at the attractor point. 

\subsubsection{Einstein equation}
The Einstein equation for the Lagrangian \eqref{lagrangian} at the attractor point is given by
\be
R_{ab}-\frac{1}{2} R \eta_{ab}= T_{ab}^{attr}
\ee
At the attractor point, the Riemann tensor is a function of spin connections (see appendix $\S$\ref{appendix}) 
which are in turn expressed in terms of the constant anholonomy coefficients. This also applies to the Ricci tensor and the scalar curvature. As a consequence the left hand side of the Einstein equation is algebraic in 
$c_{ab}^{\ \ c}$. The stress energy tensor at the attractor points is given by
\be
T_{ab}^{attr}=\V_{attr}(\phi,q)\eta_{ab} - \bigg[a_{\It \Jt}\H_{ac}^{\It}\H_{b}^{\ c \Jt}+ g^2[g_{XY} K_I^X K_J^Y +g_{\xt \yt} K_I^{\xt} K_J^{\yt}]A_a^I A_b^J \bigg].
\ee
 As one can see, the energy momentum tensor is a function of constant scalars, gauge fields and field strengths 
 at the attractor points and hence the Einstein equation reduces to an algebraic equation at the attractor points. 
 Note the appearance of the attractor potential $\V_{attr}(\phi,q)$ which is defined in \eqref{attractorpotential}. Later, 
 we show that the $\V_{attr}(\phi,q)$ follows from the scalar field equations of motion and can be constructed 
 from fermion shifts of the supersymmetry transformations.
\subsubsection{Scalar and Quaternions}
The equation of motion for the scalars $\phi^{\xt}$ in the vector and tensor multiplet is given by,
\begin{align}\label{scalareq}
\hat{e}^{-1}\d_\mu \big[\hat{e} \ g_{\zt \yt} \D^\mu \phi^{\yt} \big] &-\frac{1}{2} \frac{\d g_{\xt \yt}}{\d\phi^{\zt}}\D_\mu\phi^{\xt}\D^\mu\phi^{\yt}-g A_\mu^I g_{\xt \yt}\frac{\d K^{\xt}_I}{\d\phi^{\zt}}\D^\mu\phi^{\yt}\nn \\
&-\frac{1}{4}\frac{\d a_{\It \Jt}}{\d\phi^{\zt}}\H^{\It}_{\mu\nu} \H^{\Jt \mu\nu}-\frac{\d\V(\phi,q)}{\d\phi^{\zt}}=0.
\end{align}
Using the ansatz \eqref{ansatz1}, it can be shown that the above scalar field equation reduces to the following 
form in the tangent space:
\be\label{scalarattractor}
\frac{\d}{\d\phi^{\zt}}\bigg[\V(\phi,q)+\frac{1}{2}g^2 \ g_{\xt \yt} K^{\xt}_I K^{\yt}_J A^{Ia} A^J_a+\frac{1}{4} a_{\It \Jt} \H^{\It}_{ab}\H^{\Jt ab} \bigg]=0.
\ee
For the quaternion $q^Z$, the equation of motion is
\begin{align}\label{quaternioneq}
\hat{e}^{-1}\d_\mu \big[\hat{e}\ g_{ZY} \D^\mu q^Y \big]-\frac{1}{2}\frac{\d g_{XY}}{\d q^Z}\D_\mu q^X \D^\mu q^Y &-g A^I_\mu g_{XY} \frac{\d K^X_I}{\d q^Z} \D^\mu q^Y\nn\\ &-\frac{\d\V(\phi,q)}{\d q^Z}=0.
\end{align}
Using the ansatz\eqref{ansatz1} the quaternion equation of motion \eqref{quaternioneq} in tangent space reduces to
\be\label{quaternionattractor}
\frac{\d}{\d q^Z}\bigg[\V(\phi,q)+\frac{1}{2}g^2 g_{XY} K_I^X K_J^Y A^{aI}A^J_a \bigg]=0.
\ee
As one can see from \eqref{scalarattractor} and \eqref{quaternionattractor}, the equation of motion for the scalars at the attractor point reduces to an extremisation condition on a potential.

\subsection{Attractor Potential}
We define our attractor potential to be the one which gives rise to the attractor values of the scalars and 
quaternions upon extremization. Thus, observing the equations of motion for the scalars \eqref{scalarattractor} 
and the quaternions \eqref{quaternionattractor} the attractor potential for the $\N=2, d=5$ gauged supergravity 
can be constructed to have the form:
\be\label{attractorpotential}
\V_{attr}(\phi,q)=\bigg[\V(\phi,q)+\frac{1}{2}g^2 [\ g_{\xt \yt} K^{\xt}_I K^{\yt}_J + g_{XY} K_I^X K_J^Y] A^{Ia} A^J_a+\frac{1}{4} a_{\It \Jt} \H^{\It}_{ab}\H^{\Jt ab} \bigg]
\ee
 Note the similarity of the attractor potential \eqref{attractorpotential} with the one obtained in \cite{Kachru:2011ps} for $\N=2, d=4$ gauged supergravity. This is expected since both the theories have the same supersymmetries and the quaternionic structure. The difference is in the reality of the scalar fields and the presence of tensor fields, which contribute to the attractor potential. Thus \eqref{attractorpotential} obeys both \eqref{scalarattractor} and \eqref{quaternionattractor}.  Note that, this exact form of the attractor potential \eqref{attractorpotential} also appears in the Einstein equation. Now, we show that the potential can be constructed from fermion shifts defined at the attractor points.

In supergravity, gauging introduces additional terms in the Lagrangian that depend on the gauge coupling and 
for supersymmetry to be preserved the supersymmetry transformations have to be modified accordingly. These 
additional terms in the supersymmetry transformations can be incorporated in terms of what are called as the
 fermion shifts, which are usually defined as the non-derivative scalar dependent bosonic terms in the supersymmetry transformations of the fermions in the theory (see for eg \cite{Fre:2001jd}). 

Kachru et.al \cite{Kachru:2011ps} generalized this notion of fermionic shifts by considering the shifts at the attractor points and included terms that depend on constant gauge fields and field strengths. It was shown that the attractor potential can be written as squares of the generalized fermion shifts. In our case the generalized fermion shifts contain terms that depend on constant tensor fields as well. We will use a notation similar to that of \cite{Fre:2001jd} for defining the generalized fermion shifts.

The supersymmetry transformations \eqref{susytransformation} take the following form at the attractor points defined by \eqref{ansatz1}:
\begin{align}\label{fermionshifts1}
 \delta\psi_{ai} & = \sqrt{6} D_a\epsilon_i + (\Sigma_{i | j})^{bc} (\gamma_{abc}-4\eta_{ab}\gamma_c)\epsilon^j +\gamma_a S_{ij} \epsilon^j \nn \\
 \delta\lambda^{\at}_i &=\Sigma_{\ i | j}^{\at}\epsilon^j + (\Sigma_{\ i | j}^{\at})^a \gamma_a \epsilon^j + (\Sigma_{\ i | j}^{\at})^{ab}\gamma_{ab}\epsilon^j\nn \\
 \delta\zeta^A &=(\Sigma_{\ | j}^A) \epsilon^j + (\Sigma_{\ | j}^A)^a \gamma_a \epsilon^j
\end{align}
where, the gravitino mass matrix and shifts are given by,
\begin{align}\label{fermionshifts2}
\Sigma_{\ i | j}^{\at} &= g_R P^{\at}_{\ ij} - g W^{\at} \epsilon_{ij} &;& &(\Sigma_{\ | j}^A) &= g \N_j^{\ A}  & ; & S_{ij} &= i g_R P_{ij} ; \nn \\
(\Sigma_{\ i | j}^{\at})^a &=\frac{i}{2} g f^{\at}_{\xt}K^{\xt}_I A^{Ia}\epsilon_{ij} &;& &(\Sigma_{\ | j}^A)^a &= -\frac{i}{2} f^A_{jX} K_I^X A^{aI} & ; &  \nn\\
(\Sigma_{\ i | j}^{\at})^{ab} &=-\frac{1}{4} h^{\at}_{\It}\H^{\It ab}\epsilon_{ij} &;& &(\Sigma_{i | j})^{bc} &= -\frac{i}{4} h_{\It}\H^{bc \It}\epsilon_{ij} & ; & 
\end{align}
Using the relations \eqref{ambientmetric},\eqref{scalarvielbein}, and \eqref{quaternionvielbein} the attractor potential \eqref{attractorpotential} can be written in terms of the shifts \eqref{fermionshifts2} and their complex conjugates as follows,
\begin{align}\label{attractorpotentialshifts}
 -\V_{attr} \frac{\epsilon^l_{\ k}}{4}=  \bar{S}^i_{\ k} S_i^{\ l}-&\epsilon^{lj} \bigg\{ \big[(\overline{\Sigma^A_{ \ | k}})(\Sigma_{A | j}) +\frac{1}{2}(\overline{\Sigma^{\at i}_{\ \ | k}})(\Sigma_{\ i | j}^{\at}) \big] \nn \\
 &+\big[(\overline{\Sigma^A_{ \ | k}})_a(\Sigma_{A | j})^a +\frac{1}{2}(\overline{\Sigma^{\at i}_{\ \ | k}})_a(\Sigma_{\ i | j}^{\at})^a \big]\nn \\
 &+\big[(\overline{\Sigma^i_{\ | k}})_{ab}(\Sigma_{i | j})^{ab}+ 
(\overline{\Sigma^{\at i}_{\ \ | k}})_{ab} (\Sigma_{\ i | j}^{\at})^{ab} \big]\bigg\}
\end{align}
This relation for the attractor potential is similar to the one obtained in \cite{Kachru:2011ps} for $\N=2, d=4$ gauged supergravity. In fact, it seems that such a result could be derived for any gauged supergravity in arbitrary dimension for an attractor ansatz similar to \eqref{ansatz1}. The form of the attractor potential also makes evident the condition for the attractor to respect maximal supersymmetry. For example, the integrability condition from the gravitino supersymmetry transformation is given by,
\begin{align}\label{integrabilitycondition}
-\frac{1}{4}R_{ae}^{\ \ cd}\gamma_{cd}\epsilon_i = -\frac{1}{\sqrt{6}}&(\Sigma_{i | j})^{fc}[\omega_{a,\ f}^{\ \ b}M_{e[bc]}-\omega_{e, \ f}^{\ \ b}M_{a[bc]}]\epsilon^j\nn\\
+ \frac{1}{6}\bigg\{&[(\Sigma_{i | j})^{bc}M_{abc}+\gamma_a S_{ij}][(\Sigma_{k | l})^{gh}M_{egh}+\gamma_e S_{kl}]\nn\\
- &[(\Sigma_{i | j})^{bc}M_{ebc}+\gamma_e S_{ij}][(\Sigma_{k | l})^{gh}M_{agh}+\gamma_a S_{kl}]\bigg\}\epsilon^{jk}\epsilon^l
\end{align}
 Where $M_{abc}= \gamma_{abc}-4\eta_{ab}\gamma_c$. When all the fermionic shifts \eqref{fermionshifts2} vanish, the above equation reduces to
\be
-\frac{1}{4}R_{ae}^{\ \ cd}\gamma_{cd}\epsilon_i= \frac{1}{6} S_{ij}S_{kl}\gamma_{ae}\epsilon^{jk}\epsilon^l
\ee
The bosonic term in the right hand side is the cosmological constant, and the above equation implies Einstein equation in the absence of matter. There are no algebraic constraints on the spinors from the supersymmetry transformations of $\lambda^{\at}_i$ and $\zeta^A$ when all the fermionic shifts vanish. In such a scenario, $AdS_5$ is the only unique maximally supersymmetric ground state of this theory\cite{Ceresole:2000jd,Kallosh:2000tj}. There could also be solutions such as BPS domain walls, which preserve maximal supersymmetry. For example, In the absence of tensor multiplets, one has $W^{\at}=0$. In addition, when the vector fields vanish $\Sigma_{\ i | j}^{\at} $ and $\Sigma_{\ | j}^A$ are the only non-vanishing fermionic shifts in \eqref{fermionshifts2}. Maximal supersymmetry requires that there should be no algebraic constraints on the spinors, therefore the following terms must vanish
\be\label{maximalsusyattractorconditions}
P^{\at}_{\ ij}=h^{\at I} P_{I ij}=0 , \qquad \N_j^{\ A}=\frac{\sqrt{6}}{4}h^I K^X_I f^{A i}_X =0.
\ee
The above equations lead to the attractor conditions derived in \cite{Ceresole:2001wi} for domain wall solutions that interpolate between AdS vacua. Such planar domain wall solutions characterised by constant anholonomy coefficients are non-trivial examples of supersymmetric generalised attractors.

For non-supersymmetric attractors or attractors that preserve a part of the supersymmetry there are non-vanishing shifts. Hence there will be constraints on the spinors, as a result of which one will either have some amount of supersymmetry preserved (which is expected at least for Lifshitz solutions \cite{Cassani:2011sv}) or none at all. In the cases where one deals only with vector multiplets, the integrability conditions on the Killing spinors have been worked out in \cite{Gauntlett:2003fk} and the constraints imply that one gets either $1/2$ BPS or $1/4$ BPS solutions. We later work out some simple examples of Bianchi attractors \cite{Iizuka:2012iv} from gauged supergravity with only vector multiplets, we expect these solutions to be $1/4$ BPS.

\section{Constant Anholonomy and Homogeneity}\label{constantanholonomy}
\subsection{Constant anholonomy:}
The assumption of constant anholonomy, though restrictive is sufficiently general to capture the near horizon geometries of extremal black branes. Several examples such as $dS_4$, $AdS_4$, Lifshitz and Schrodinger geometries are known to have constant anholonomy\cite{Kachru:2011ps}. In this section, we take a generic form of an extremal black brane metric that belongs to the Bianchi type I (see \cite{Iizuka:2012iv}), impose constant anholonomy and determine the restrictions it puts on the form of the metric. 

Let us consider a black brane metric of the form:
\be\label{blackbranemetric}
ds^2= -a(r)^2 dt^2+ \frac{dr^2}{b(r)^2}+c(r)^2 dx^2+d(r)^2 dy^2+ e(r)^2 dz^2\ ,
\ee
where $a(r),b(r),c(r),d(r)$ and $e(r)$  are all functions of $r$. The other Bianchi types are off-diagonal in 
$(x,y,z)$ directions and the metric contains functions dependent on these variables. We consider this simple case to illustrate the basic idea. The f\"{u}nfbeins for 
the metric are
\be
e^0_t=a(r) ,\ e^1_r=\frac{1}{b(r)}, \ e^2_x = c(r), \ e^3_y= d(r), \ e^4_z= e(r).
\ee
The only independent non-vanishing anholonomy coefficients \eqref{anholonomy} are ,
\begin{align}
 c_{01}^{\ \ 0}  &= b(r)\frac{a'(r)}{a(r)}, &c_{21}^{\ \ 2} &= b(r)\frac{c'(r)}{c(r)} , &c_{31}^{\ \ 3} &= b(r)\frac{d'(r)}{d(r)} , &c_{41}^{\ \ 4} &= b(r)\frac{e'(r)}{e(r)} \ ,
\end{align}
where the prime indicates derivative w.r.t $r$. Demanding constant anholonomy coefficients leads to the following equations:
\begin{align} \label{metricconstraints}
\frac{a'(r)}{a(r)} =\frac{C_0}{b(r)} ,\ \ \frac{c'(r)}{c(r)} =\frac{C_2}{b(r)} ,\ \ \frac{d'(r)}{d(r)} =\frac{C_3}{b(r)},\ \ \frac{e'(r)}{e(r)} =\frac{C_4}{b(r)},
\end{align}
where $C_0, C_2 ,C_3 , C_4$ are the constant values of the anholonomy coefficients. Since we have assumed all the unknown functions to be pure functions of $r$, we may treat the above partial differential equations as ordinary differential equations. 
We consider some specific cases. The first case,  $ b(r)=a(r)$ leads to the near horizon geometry of the extremal AdS Reissner-Nordstrom black hole and the second case, $ b(r)=c(r)$ gives rise to the Lifshitz geometry.

\, \emph{case} $i)\ b(r)=a(r)$ :
The metric takes the following form,
\be\label{NHEblackbranemetric}
ds^2=-C_0^2 \rt^2 dt^2+ \frac{d \rt^2}{C_0^2 \rt^2}+ \rt^{2\frac{C_2}{C_0}} d\xt^2+ \rt^{2\frac{C_3}{C_0}} d\yt^2+ \rt^{2\frac{C_4}{C_0}} d\zt^2 .
\ee
where $\rt=r+ \frac{a_0}{C_0}$ and $(\xt,\yt,\zt)= (a_2 x,a_3 y, a_4 z)$. Here all the $a_i$ are integration constants. The metric \eqref{NHEblackbranemetric} is the near horizon geometry of the extremal Reissner-Nordstrom black brane with the identifications $t=\tt,\ C_0=\sqrt{\Lambda}, C_2=C_3=C_4=0 $ and $a_2=a_3=a_4=r_h$.

\, \emph{case} $ii)\ b(r)=c(r)$ :
Solving for the other functions, the metric takes the form
\be\label{lifshitzmetric}
ds^2=- \rt^{2\frac{C_0}{C_2}}d\tt^2+ \frac{d \rt^2}{C_2^2 \rt^2}+ \rt^2 C_2^2 dx^2+ \rt^{2\frac{C_3}{C_2}} d\yt^2+ \rt^{2\frac{C_4}{C_2}} d\zt^2 .
\ee
where $\rt=r+ \frac{a_2}{C_2}$ and $(\tt,\yt,\zt)= (a_0 t,a_3 y, a_4 z)$. The $a_i$'s are all integration constants. 
One can identify \eqref{lifshitzmetric} with the Lifshitz metric by the identifications $C_0=\frac{u}{L} , C_2=C_3=C_4=\frac{1}{L}$ and $a_0=\frac{1}{L^u}, a_3=a_4=
\frac{1}{L}$, where $L$ is the size of the spacetime. Redefining $\hat{t}=L t, \ (\hat{r},\hat{x},\hat{y},\hat{z})= \frac{1}{L}(\rt,x,y,z)$, one gets the standard Lifshitz metric (see for eg \cite{Cassani:2011sv})
\be\label{lifshitzmetricstandard}
ds^2=L^2\bigg[-\hat{r}^{2u} d\hat{t}^2+ \frac{d \hat{r}^2} {\hat{r}^2}+ \hat{r}^2(d\hat{x}^2+d\hat{y}^2+d\hat{z}^2)\bigg] .
\ee
Similarly the Anisotropic Lifshitz \cite{Taylor:2008tg,Pal:2009yp} metric can be identified with \eqref{lifshitzmetric} by  choosing $C_0=\frac{u}{L}, C_2=\frac{1}{L}, C_3=\frac{v}{L}, C_4=\frac{w}{L}$ and $a_0=\frac{1}{L^u}, a_3=\frac{1}{L^v}, a_4=\frac{1}{L^w}$:
\be
ds^2=L^2\bigg[-\hat{r}^{2u} d\hat{t}^2+ \frac{d \hat{r}^2} {\hat{r}^2}+ \hat{r}^2 d\hat{x}^2+ \hat{r}^{2v} d\hat{y}^2+ \hat{r}^{2w} d\hat{z}^2\bigg].
\ee
Thus, one can see that constant anholonomy requires the extremal black brane metric \eqref{blackbranemetric} to have a specific form such as \eqref{NHEblackbranemetric} or \eqref{lifshitzmetric}. 

The assumption of constant anholonomy itself is an implied symmetry of the metric \eqref{blackbranemetric}. The hypersurfaces on which the algebra of vielbeins $\et_a$ have constant anholonomy coefficients are called surfaces of transitivity and the vectors $\et_a$ generate a simply transitive group. It is known that for homogeneous spacetimes with spacelike hypersurfaces of dimension three there exists Lie groups of symmetries that act simply transitively on the surfaces\cite{wald1984general}. Thus the algebra of the invariant vectors \eqref{anholonomy} can be shown to be isomorphic to the real Lie algebras of dimension three, which were classified by Bianchi\cite{1975classical}. The Bianchi classification is well known in cosmology and was used recently to classify extremal black branes with spatially homogeneous directions \cite{Iizuka:2012iv}. We briefly explain the connection between constant anholonomy coefficients and homogeneous spaces.

Consider a basis of Killing vectors that generate a simply transitive group of dimension three. These Killing vectors have the algebra,
\be
[\xi_\mu,\xi_\nu]= \tilde{C}_{\mu\nu}^{\ \ \lambda}\xi_\lambda.
\ee
For each of the Bianchi classes, one can go to a suitable basis and construct invariant vector fields $\et_a$ that commute with the Killing vectors,
\be\label{killingcommute}
[\xi_\mu,\et_a]=0.
\ee
Now, the Jacobi identity between $(\et_a,\xi_\mu,\xi_\nu)$ implies $\tilde{C}_{\mu\nu}^{\ \ \lambda}$ are constants in spacetime. These are the structure constants of the three dimensional real Lie algebras given by the Bianchi classification. The Jacobi identity between $(\et_b,\et_a,\xi_\mu)$ together with \eqref{killingcommute} imply that the anholonomy coefficients $c_{ab}^{\ \ c}$ are constants in the surface of transitivity. 

Alternatively, given that the invariant one forms have an algebra \eqref{anholonomy} with constant anholonomy coefficients, \cite{Ellis} have shown that \eqref{killingcommute} is satisfied by three independent Killing vectors provided the following conditions are satisfied:
\be\label{conditiononanholonomy}
c_{0a}^{\ \ 0}= c_{ab}^{\ \ 0}=0.
\ee
A quick look at the metric \eqref{blackbranemetric}, its vielbeins and non-vanishing anholonomy coefficients shows that both the conditions hold good for all $a,b= 2,3,4$. This implies \eqref{killingcommute} is satisfied for the spatial directions $(x,y,z)$, which means that these directions are homogeneous. Note that the conditions \eqref{conditiononanholonomy} are also satisfied for all the Bianchi type metrics (see Appendix A of \cite{Iizuka:2012iv}).

 One could also consider situations when there are more homogeneous directions (e.g. 
homogeneity along the time direction). The symmetry groups of such metrics might as well be classified by analogues of the Bianchi classification in higher dimensions. In mathematics literature there exists classifications of real Lie algebras in higher dimensions \cite{patera:986}. Similar to the three dimensional case, one could ask if the algebra of the invariant one forms is isomorphic to these real Lie algebras. However, It turns out that a general proof for this is not known for 
dimensions greater than three. Nevertheless, there has been some work in the physics literature where such cases have been considered \cite{Hervik:2002sn,Hervik:2002af}. 

\subsection{Some examples:}

In this section, we show that some of the simple Bianchi type metrics can be realised from simple $d=5$ gauged supergravity models. Our objective here is not to be exhaustive regarding the possibilities, as this has already been considered in \cite{Iizuka:2012iv}. We will take some of the explicit examples considered in \cite{Iizuka:2012iv} and show that they can be obtained from a specific gauged supergravity model. We are motivated by the observation that most of the Bianchi attractors constructed in \cite{Iizuka:2012iv} are sourced by massive gauge fields. In gauged supergravity one does not have explicit massive gauge fields, as these would break supersymmetry. Nevertheless, expanding the scalar kinetic term one gets terms like $g^2 g_{\xt\yt} K^{\xt}_I K^{\yt}_J A^I_aA^{J a}$ that are proportional to square of the gauge field. These terms appear due to the presence of covariant derivatives which appear due to the gauging. Since the scalars are constant at the attractor point, the coefficients of these 
terms act like a mass for the gauge field. Hence, one can expect to realise the Bianchi attractors from specific truncations of gauged 
supergravity models.

We will consider a specific gauged supergravity model constructed by \cite{Gunaydin:2000xk} and construct some of the simplest Bianchi type solutions. This model is a truncation of the general $\N=2$ gauged supergravity in 5 dimensions \cite{Ceresole:2000jd}. It has a vector multiplet (There is also the graviphoton), two tensor multiplets and no hypermultiplets. The three real scalars in the theory parametrise a manifold $\SS$ with the symmetry 
group
\be
G=SO(1,1) \times \frac{SO(2,1)}{SO(2)}.
\ee
In \cite{Gunaydin:2000xk}, the two possible gaugings $SO(2)$ and $SO(1,1)$ of the symmetries of $\SS$ together with gauging of $U(1)_R$ symmetry\footnote{When $n_H=0$ the $SU(2)_R$ symmetry of the general gauged supergravity \cite{Ceresole:2000jd} is broken to $U(1)_R$.} were considered and the critical points of the full potential were analysed in both cases. 

We consider the gauging of the $SO(2)$ subgroup of $G$ for our purpose\footnote{The non-compact $SO(1,1)$ gauging has only one critical point corresponding to a deSitter vacuum and breaks $\N=2$ supersymmetry.}. The group $SO(2)$ has only one generator and the Killing vector field that generates this symmetry is given by
\be\label{killingvectorSO2}
K_0^{\xt}=\bigg\{ -\frac{\phi^1}{||\phi||^2}, \frac{\phi^2}{||\phi||^2}, \frac{\phi^3}{||\phi||^2} \bigg\}.
\ee
The metric on the scalar manifold $g_{\xt \yt}$ is given in $\S$ \ref{appendix1}. The graviphoton $A^0$ is chosen to be the $SO(2)$ gauge field.
In addition $A[U(1)_R]=A^0 V_0 + A^1 V_1$ is chosen to be the $U(1)_R$ gauge field. The parameters $V_0$ and $V_1$ are free\footnote{For a general non abelian $K$, the $V_I$ satisfy $V_I f^I_{ \ JK}=0$. For abelian $K$, the structure constants vanish.}, but satisfy constraints determined by the critical points of the potential
\be\label{scalarpotentialSO2}
\V(\phi^1,\phi^2,\phi^3)= \frac{g^2}{8} \bigg[ \frac{[(\phi^2)^2+(\phi^3)^2]}{||\phi||^6}\bigg]-2 g_R^2 \bigg[2\sqrt{2} \frac{\phi^1}{||\phi||^2} V_0 V_1 + ||\phi||^2 V_1^2\bigg].
\ee
We refer the reader to \cite{Gunaydin:2000xk} for a detailed analysis of the critical points of the potential. Some of the details required for the computation are summarised in the appendix $\S$ \ref{appendix1}. We will consider only the critical points that have AdS vaccum and $\N=2$ supersymmetry\footnote{For pure $U(1)_R$ gauging the supersymmetric critical point is always a maximum. The potential due to $SO(2)$ gauging has a $g^2$ dependent term that makes this a supersymmetric saddle point.}. One motivation for this is due to the result of \cite{Cassani:2011sv}, where it was shown that $d$ dimensional Lifshitz solutions of scaling dimension 2 exist at extrema of $d+1$ dimensional potentials with negative values, i.e at AdS vaccum. Another point is that the equations of motion (discussed in $\S$ \ref{eom}) in component form simplify significantly when evaluated at this particular saddle point. The conditions for $\N=2$ supersymmetry and AdS vaccum are given by extremising \eqref{scalarpotentialSO2},
\be\label{v0v1constraintforAdSN=2}
\phi^2=0, \quad \phi^3=0,\quad \phi^1_c=\bigg(\sqrt{2}\frac{V_0}{V_1}\bigg)^{\frac{1}{3}}, \quad V_0 V_1 > 0 , \quad 32 \frac{g_R^2}{g^2}V_0^2\leq1.
\ee
The value of the potential \eqref{scalarpotentialSO2} evaluated at \eqref{v0v1constraintforAdSN=2} is the AdS cosmological constant $\V_{AdS}= -6 g_R^2(\phi^1_c)^2 V_1^2$. 

In the following subsections we construct a Lifshitz solution with scaling dimension 3, a Bianchi type II and a Bianchi type VI solution of the gauged supergravity \cite{Gunaydin:2000xk}. In all these solutions the scalars take values for which the theory would have AdS vacuum.

\subsubsection{Bianchi type I: Lifshitz solution}\label{Lifshitz3solution}
We now look for Lifshitz like solutions within this model. We take the metric ansatz to be of the form given in \eqref{lifshitzmetricstandard} and solve the equations of motion derived in $\S$ \ref{eom} for this theory. The Lifshitz solutions considered in the literature are often sourced by massive time like gauge fields \cite{Cassani:2011sv,Halmagyi:2011xh}. We assume that the $SO(2)$ gauge field\footnote{ We have used the notation $A^{Ia}$ earlier with $I$ labelling the vectors and $a$ the tangent space index. In component form, to avoid confusion we will denote the tangent space indices with with an overbar, i.e $A^{0\bar{0}}$} has only time like component 
\be\label{gaugefieldansatz}
A^{0t}=e^t_{\bar{0}} A^{0\bar{0}}=\frac{\ \ \hat{r}^{-u} }{L} A^{0\bar{0}}
\ee
We do not make any assumptions on the other gauge field $A^1$. We take it to be of the general form $A^{1\mu}=e^\mu_a A^{1 a}$.
As a further simplification we also assume that all the tensor field components vanish. This need not be true for a more general theory with different gauging or a different metric ansatz. However, we find  that the Lifshitz like ansatz does not admit any consistent solution with non-vanishing tensor fields within the model considered. We will explain the reason for this towards the end of the section.

As before, we will work in tangent space. The undetermined parameters are $A^{0\bar{0}}, A^{1\bar{0}}, A^{1\bar{1}},A^{1\bar{2}},A^{1\bar{3}},A^{1\bar{4}}$, the scaling parameter $u$, size of the spacetime $L$. These are to be determined in terms of the gauge couplings $g, g_R$ and the free parameters $V_0, V_1$ which are constrained by \eqref{v0v1constraintforAdSN=2}.

The equations \eqref{gaugefieldeq} for the gauge fields $A^0$ and $A^1$ evaluated at the critical point \eqref{v0v1constraintforAdSN=2} read as follows:
\begin{align}\label{gaugefieldeqSO2}
 \hat{e} A^{0\bar{0}} (g^2 L^2 - u (\phi^1_c)^8) &= 0 \nn\\
 \hat{e} A^{1\bar{0}} u &=0 \nn \\
 \hat{e} A^{1\bar{2}} (2+u) &=0\nn\\
 \hat{e} A^{1\bar{3}} (2+u) &=0\nn\\
 \hat{e} A^{1\bar{4}} (2+u) &=0 \ , 
\end{align}
whereas the off-diagonal components of the Einstein field equations are,
\begin{align}\label{einsteineqS02}
 A^{1\bar{0}}A^{1\bar{2}} u &=0\nn\\
 A^{1\bar{0}}A^{1\bar{3}} u &=0\nn\\
 A^{1\bar{0}}A^{1\bar{4}} u &=0\nn\\
 A^{1\bar{2}}A^{1\bar{3}} &=0 \nn\\
 A^{1\bar{2}}A^{1\bar{4}} &=0\nn\\
 A^{1\bar{3}}A^{1\bar{4}} &=0.
\end{align}
The gauge field equations of motion \eqref{gaugefieldeqSO2} imply that $A^{1\bar{0}}=0$ for a non-zero $u$. The off-diagonal Einstein equations imply that any two of the three components $A^{1\bar{2}},A^{1\bar{3}},A^{1\bar{4}}$ must vanish. If we take say $A^{1\bar{3}}=A^{1\bar{4}}=0 $, then the gauge field equation for $A^{1\bar{2}}$ give $u=-2$ which is inconsistent with the equation for $A^{0\bar{0}}$. Hence, we set all three of them to zero. 
Note that this still leaves $A^{1\bar{1}}$ unfixed\footnote{For all the Bianchi classes the Field strengths do not depend upon $A^{1\bar{1}}$, so this component can enter only through an $A^{1 a} A^1_a$ term or the Chern-Simons term. The former does not happen here as $A^{1a}$ is not used to gauge the symmetries of the scalar manifold. The latter does not occur since topological terms do not contribute in this case.}. With these simplifications, the diagonal $(tt,rr,xx)$ components of the Einstein equation are, 
\begin{align}
 12 (\phi^1_c)^4+ (A^{0\bar{0}})^2 (3 g^2 L^2+u^2 (\phi^1_c)^8)-24 L^2 g_R^2 V_0^2 &=0\label{eq3}\\
 -6(1+u)(\phi^1_c)^4+(A^{0\bar{0}})^2 (3 g^2 L^2-u^2 (\phi^1_c)^8)+24 L^2 g_R^2 V_0^2 &=0\label{eq4}\\
-2(3+2u+u^2)(\phi^1_c)^4+(A^{0\bar{0}})^2 (3 g^2 L^2+u^2 (\phi^1_c)^8)+24 L^2 g_R^2 V_0^2 &=0\label{eq5}
\end{align}
The $(yy,zz)$ components give the same equations as the $xx$ one. Subtracting \eqref{eq4} and \eqref{eq5}, $A^{0\bar{0}}$ can be determined as,
\be\label{SO2gaugefield}
A^{0\bar{0}}= \sqrt{\frac{u-1}{u}}\frac{1}{(\phi^1_c)^2} \ , 
\ee
where we have chosen the positive sign for $A^{0\bar{0}}$. The values of $L$ and $u$ can be determined from the gauge field and scalar field equations. Substituting $A^{1\bar{2}}=A^{1\bar{3}}=A^{1\bar{4}}=0$, the gauge field equations \eqref{gaugefieldeqSO2} reduce to 
\be\label{eq1}
g^2 L^2 - u (\phi^1_c)^8=0.
\ee
The scalar field equations\eqref{scalareq} evaluated at the attractor point\eqref{v0v1constraintforAdSN=2} must vanish and this gives the relation
\be\label{eq2}
3 g^2 L^2 -u^2 (\phi^1_c)^8=0.
\ee
The two equations \eqref{eq1} and \eqref{eq2} can be solved to get,
\be\label{exponentsize}
u=3; \quad L= \sqrt{3} \frac{(\phi^1_c)^4}{g}
\ee
Substituting the values of \eqref{exponentsize}, \eqref{SO2gaugefield} in \eqref{eq3}, one gets the following constraint that relates the free parameters $V_0,V_1$ to the ratio of the couplings $g$ and $g_R$.
\be\label{constraint}
\frac{1}{3 (\phi^1_c)^4}=\frac{g_R^2}{g^2}V_0^2
\ee
Let us summarise the solution,\\
\fbox{
\begin{minipage}{\linewidth}
\begin{align}\label{solution}
 & ds^2 = L^2\bigg[-\hat{r}^{2u} d\hat{t}^2+ \frac{d \hat{r}^2} {\hat{r}^2}+ \hat{r}^2(d\hat{x}^2+d\hat{y}^2+d\hat{z}^2)\bigg]\nn\\
 & u=3; \quad L= \sqrt{3} \frac{(\phi^1_c)^4}{g}; \quad A^{0\bar{0}}=\sqrt{\frac{2}{3}}\frac{1}{(\phi^1_c)^2} \nn\\
 & \phi^1_c=\bigg(\sqrt{2}\frac{V_0}{V_1}\bigg)^{\frac{1}{3}}; \quad V_0 V_1 > 0 ; \quad\frac{32}{3 (\phi^1_c)^4}\leq1.
\end{align}
\end{minipage}
}
The attractor potential for the above solution is given by, 
\be\label{attractorpotentialLifshitz}
\V_{attr}(\phi^1_c)= - \bigg[\frac{(A^{0\bar{0}})^2}{2(\phi^1_c)^4} \bigg(3g^2+ u^2 \frac{(\phi^1_c)^8}{L^2}\bigg)-\V_{AdS}\bigg]
\ee
where $\V_{AdS}= -6 g_R^2(\phi^1_c)^2 V_1^2$ is the cosmological constant. 

The attractor potential can be written in terms of the fermionic shifts \eqref{fermionshifts2} defined earlier. The shifts $(\Sigma_{\ | j}^A)^a,\Sigma_{\ | j}^A $ vanish since there are no hypermultiplets in the theory. The shift $\Sigma_{\ i | j}^{\at}$ vanishes due to the choice \eqref{v0v1constraintforAdSN=2}\footnote{$P^{\at}_{ij}, W^{\at}$ vanish for this choice \cite{Gunaydin:2000xk}.}. The remaining shifts are non-vanishing at the attractor points. Since some of the shifts are non-vanishing, the solution \eqref{solution} preserves only a part of the supersymmetry. In fact it is known that Lifshitz solutions from $\N=2$ supergravities preserve $1/4$ of the supersymmetry\cite{Cassani:2011sv}.

\subsubsection{Some other Bianchi type metrics}
In this subsection we give some more examples. In particular, the Bianchi type II, VI attractors arising from gauged supergravity. The analysis is entirely parallel to the previous section. Hence we briefly summarise the solutions and give the necessary equations in the appendix $\S$\ref{bianchi26fieldeq}. We refer the reader to \cite{Ellis,ryan1975homogeneous,Iizuka:2012iv} for details on the various Bianchi classes, invariant forms and metrics. As before, we consider only the time like component for the $SO(2)$ gauge field \eqref{gaugefieldansatz} to be non-vanishing and set all the tensor fields to be zero. We also find that the off-diagonal Einstein equations for all cases imply $A^{1\bar{0}}=A^{1\bar{2}}=A^{1\bar{3}}=A^{1\bar{4}}=0$ while $A^{1\bar{1}}$ is left unfixed. The rest of the equations are solved in the appendix for the type II and type VI cases. 

The type II solution is given by\\
\fbox{
\begin{minipage}{\linewidth}
\begin{align}\label{solutionII}
 & ds^2 = L^2\bigg[-\hat{r}^{2u} d\hat{t}^2+ \frac{d \hat{r}^2} {\hat{r}^2}+ \hat{r}^{2w}d\hat{x}^2+ \hat{r}^{2(v+w)} d\hat{y}^2\nn\\
& \quad\quad\quad\quad\quad\quad-2\hat{x} \hat{r}^{2(v+w)} d\hat{y} d\hat{x} +[\hat{r}^{2(v+w)}\hat{x}^2+\hat{r}^{2v}] d\hat{z}^2 \bigg]\nn\\
  &u=\sqrt{2};\quad v=w=\frac{1}{2\sqrt{2}};\quad L= \sqrt{\frac{2}{3}} \frac{(\phi^1_c)^4}{g}; \quad A^{0\bar{0}}=\sqrt{\frac{5}{8}}\frac{1}{(\phi^1_c)^2}; \nn\\
  &\phi^1_c=\bigg(\sqrt{2}\frac{V_0}{V_1}\bigg)^{\frac{1}{3}}; \quad V_0 V_1 > 0 ; \quad\frac{23}{2 (\phi^1_c)^4}\leq1.
\end{align}
\end{minipage}
}\\{}\\
whereas the type VI solution is given by,\\{}\\
\fbox{
\begin{minipage}{\linewidth}
\begin{align}\label{solutionVI}
 & ds^2 = L^2\bigg[-\hat{r}^{2u} d\hat{t}^2+ \frac{d \hat{r}^2} {\hat{r}^2}+ d\hat{x}^2 + e^{-2\hat{x}} \hat{r}^{2v} d\hat{y}^2+e^{-2h\hat{x}} \hat{r}^{2w} d\hat{z}^2\bigg]\nn\\
& u=\frac{1}{\sqrt{2}}(1-h);\quad v=-\frac{1}{\sqrt{2}}h;\quad w=\frac{1}{\sqrt{2}};\quad L= \frac{(\phi_1^c)^4}{\sqrt{6}g}(1-h);\quad \nn\\
& A^{0\bar{0}}=\sqrt{\frac{-2h}{(-1+h)^2}}\frac{1}{(\phi^1_c)^2};\quad h<0; \quad h\neq0,1; \nn\\
&\phi^1_c=\bigg(\sqrt{2}\frac{V_0}{V_1}\bigg)^{\frac{1}{3}}; \quad V_0 V_1 > 0; \quad \frac{8(3-h+3h^2)}{(\phi^1_c)^4(-1+h)^2}\leq 1
\end{align}

\end{minipage}
}\\{}\\
As one can see from the above equations, we require $h<0$ for the gauge field to be real which agrees with \cite{Iizuka:2012iv}. In deriving this particular solution, we also required in addition $h\neq0,1$. These two cases correspond to the Bianchi type III and type V metrics which can be realised as limiting cases of type VI. The type V metric is obtained in the $h\rightarrow 1$ limit of the type VI metric. In \cite{Iizuka:2012iv} it was found that the solution exists in the massless limit. In this case the equivalent of a massless limit would be to take $g\rightarrow 0$ as $h\rightarrow 1$. Even though the length of the space time can be kept finite the time component of the gauge field blows up in this limit. Thus in this model we cannot obtain the type V solution in this manner. A similar issue occurs for the $h\rightarrow 0$ limit for the type III metric. In this case, the gauge field vanishes. In both situations one cannot take either $V_1$ or $V_0$ to zero, as this would jeopardise the gauging 
procedure. In summary, the type V and type III metrics do not seem to be valid attractors of the gauged supergravity considered here. However, they may still be solutions to some generic supergravity that belongs to the same class. For example, the type VII metric requires two massive gauge fields to start with, therefore one has to start from a supergravity model that uses two gauge fields to gauge the symmetries of the scalar manifold. Such gauged supergravities may be constructed based on the generic Jordan class of scalar manifolds. We refer the reader to  \cite{Gunaydin:1999zx,Gunaydin1985573,Gunaydin1984244} for more details on such supergravity theories.

The attractor potential for the cases considered here is the same as \eqref{attractorpotentialLifshitz} with the values of the parameters and constraints specific to each case. The solutions are determined by the parameters $g, V_0 $ and $V_1$ together with the constraints. In this section, We have given a general idea of how to get such metrics from a simple gauged supergravity model via the generalised attractor ansatz \eqref{ansatz1}. The other Bianchi classes may be realised in a similar way from more generic gauged supergravities.

We now conclude this section with a few comments. Let us first note that the Chern-Simons term had no contribution whatsoever for any of these solutions. In particular as observed in \cite{Cassani:2011sv}, topological terms vanish for the Lifshitz like solution sourced by a time-like gauge field. Remember that the field strengths are written in terms of the anholonomy coefficients. For the Lifshitz like solution and in general for any Bianchi type I metric the non-vanishing anholonomy coefficients are $c_{01}^{\ \ 0},c_{21}^{\ \ 2},c_{31}^{\ \ 3},c_{41}^{\ \ 4}$. Due to this the Chern-Simons term $\epsilon^{bacde}c_{ba}^{\ \ f} c_{cd}^{\ \ g} A_f A_g A_e$ vanishes. For similar reasons the structure constants of the Bianchi classes \cite{ryan1975homogeneous} imply that there can be no support from the Chern-Simons term for any of the Bianchi type metrics which are sourced by time-like (or space-like) gauge fields. Note that for metrics with homogeneous directions greater than three, if the possible symmetry 
groups are given by the classification of real Lie algebras (see, for example,  table I of \cite{patera:986}), the topological terms could have an effect on the solution. 

Another important point to discuss here is the absence of tensor fields. In the literature there are known anisotropic Lifshitz solutions sourced by massive two forms \cite{Taylor:2008tg}. However, it is not possible to realise such solutions within gauged supergravity. Unlike in \cite{Taylor:2008tg}, the kinetic terms for the tensor fields in gauged supergravity have a toplogical origin \eqref{lagrangian}. In fact, the kinetic term for the tensor field comes from the Chern-Simons term in the original ungauged supergravity. Therefore, we do not expect the tensor fields in the theory to contribute to Lifshitz like metrics. In the supergravity model under consideration we have verified that the tensor fields do not contribute to the other Bianchi type metrics. This is in accordance with the results of \cite{Iizuka:2012iv} where such metrics were supported only by the gauge fields.

\section{Summary}\label{summary}

We studied the generalised attractors in $\N=2,d=5$ gauged supergravity defined by constant anholonomy, constant gauge fields, constant tensor fields and constant scalars at the attractor points. We showed that all the equations of motion become algebraic at the attractor points. We constructed the attractor potential from the scalar field equations and showed that it can be written independently from squares of the bosonic terms in the fermion supersymmetry transformations. We argued that all the attractors of this theory would be either partly supersymmetric or non-supersymmetric based on the killing spinor integrability conditions. We showed that some of the simplest Bianchi attractors sourced by massive gauge fields can be realised from gauged supergravity models. In particular, we constructed a Lifshitz solution with scaling dimension 3, a Bianchi type II and a Bianchi type VI solution from the gauged supergravity model of \cite{Gunaydin:2000xk}.

The analysis considered in this paper together with that of \cite{Kachru:2011ps} suggests that a similar analysis can be performed for other gauged supergravity theories with different supersymmetries such as $\N=4, d=5$ \cite{Schon:2006kz,Dall'Agata:2001vb} and $\N=3, d=4$ gauged supergravity \cite{Castellani:1985ka}.  In another note, for blackhole solutions in ungauged supergravity the attractor mechanism can be equivalently understood from the entropy function formalism \cite{Sen:2007qy}. It will be useful to explore a similar understanding for the generalised attractors in gauged supergravity. 

Although in this paper we have restricted ourselves to abelian gauging, the analysis should equally apply for non-abelian gaugings. Of course, one has to replace the abelian field strength $F_{\mu\nu}^{\ \ I}$ with its non-abelian counterpart and therefore, the equations of motion will have additional terms. However because of the ansatz \eqref{ansatz1}, one will still get algebraic equations in tangent space and the attractor potential can be constructed as before from fermionic shifts. The important differences are, in addition to the requirements of non-abelian gaugings \cite{Andrianopoli:1996cm}, the parameters $V_I$ used in the gauging of R-symmetries are constrained by $V_I f^{I}_{\ JK} =0$; where $f^{I}_{\ JK}$ are structure constants of the gauge group $K$. Furthermore, in the absence of tensor multiplets the $C_{IJK}$ also satisfy a similar constraint \cite{Ceresole:2001wi}. It would be very interesting to construct explicit examples and study the case in detail.

As we have already pointed out, it would be interesting to consider supergravity models with more vector 
multiplets to see if we can embed other Bianchi type solutions. It would be more interesting to consider 
examples of generalised attractors in gauged supergravity models with non-trivial tensor fields and study
their relevance in more detail. At this point it is also natural to ask if we can find a string theory embedding 
of these simple gauged supergravity models by suitably restricting the tensor $C_{IJK}$. Similarly it is 
worth investigating the $4d \rightarrow 5d$ lift along the line of \cite{Gaiotto:2005xt}. Finally, it would be 
interesting to study the CFT duals of the examples considered here. We hope to investigate some of 
the issues raised here in future.

\noindent
\textbf{Acknowledgements}: We would like to thank Sandip Trivedi for helpful discussions. K.I would like
to  thank Swastik Bhattacharya, Sudipto Paul Chowdhury, Rohan Poojary and Nilanjan Sircar for useful 
discussions. K.I also thanks the organisers of Asian winter school, Japan for giving the opportunity to present a preliminary version of this work as well as the participants of ICTP spring school for several stimulating discussions. The work of K.I is supported by a research fellowship from the Institute of Mathematical Sciences, Chennai. The work of PKT is partially supported by IFCPAR/CEFIPRA project No 4204-2.

\appendix
\section{Notations and Conventions:}\label{appendix}
For most of the paper we use the conventions of \cite{Ceresole:2000jd}. We summarise them here for the convenience of the reader.
\begin{itemize}
 \item Greek indices $\mu, \nu,\ldots$ denote space time indices with $\mu=0,1,\ldots,4$. The space time metric is $g_{\mu\nu}$.
 \item Latin indices $a,b,\ldots$ denote tangent space indices with $a=0,1\ldots,4$. The tangent space metric has the signature $\eta_{ab}= \{-,+,+,+,+\}$
 \item For the symplectic majorana spinors in the theory $i,j$ are used to denote the $USp(2)$ indices with $i=1,2$. The $USp(2)$ indices are raised and lowered by 
\be
A^i=\epsilon^{ij}A_j,\quad A_i=A^j\epsilon_{ji}
\ee
with $\epsilon_{12}=\epsilon^{12}=1$. The advantage with this choice is the covariance of reality relations and $\epsilon$ contractions. However, the mixed $\epsilon$ tensors are antisymmetric.
\be
\epsilon^{jk}\epsilon_{ki}=\epsilon^{j}_{\ i}=-\delta^j_i=-\epsilon^{\ j}_{i}
\ee
\end{itemize}
\subsection{Anholonomy coefficients:}
The f\"{u}nfbein $e^a_\mu(x)$ are related to the space time metric by
\be
g_{\mu\nu}=e^a_\mu e^b_\nu \eta^{ab}.
\ee
Defining the one form $e^a\equiv e^a_\mu dx^\mu$ and its dual $ \et_a \equiv e^\mu_a \d_\mu$, the anholonomy coefficients are defined as Lie brackets of the dual f\"{u}nfbein;
\be\label{anholonomy} 
[\et_a,\et_b] \equiv c_{ab}^{\ \ c} \et_c ; \quad  c_{ab}^{\ \ c} = e^\mu_a e^\nu_b (\d_\nu e_\mu^c -\d_\mu e_\nu^c)
\ee
The tangent space curvature can be written in terms of the anholonomy coefficients and the spin connection\footnote{The tangent space covariant derivative is defined as $D_a V^b = \d_a V^b + \omega_{a, \ c}^{\ \ b} V^c$} (see for eg \cite{wald1984general,ortin2007gravity})
\be\label{riemann}
R_{abc}^{\ \ \ d}= \d_a\omega_{bc}^{\ \ d}-\d_b\omega_{ac}^{\ \ d}-\omega_{ac}^{\ \ e}\omega_{be}^{\ \ d}+\omega_{bc}^{\ \ e}\omega_{ae}^{\ \ d}-c_{ab}^{\ \ e}\omega_{ec}^{\ \ d}
\ee
In the absence of torsion, as in the case of Riemann spacetime, the spin connection and anholonomy coefficients are related as follows:
\be\label{anholonomyandspinconnection}
\omega_{a,bc} =\frac{1}{2} [ c_{ab,c}- c_{ac,b}- c_{bc,a}], 
\ee
where $\omega_{a ,bc}=-\omega_{a ,cb} $ and $c_{ab,c}=- c_{ba,c}$. It follows that when one takes constant $c_{ab}^{\ \ c}$, the derivatives in \eqref{riemann} vanish and the Riemann tensor is just a function of the constant anholonomy coefficients and is non singular. This also applies to the Ricci tensor and scalar curvature, as a consequence the left hand side of the Einstein equation is algebraic in $c_{ab}^{\ \ c}$. Another consequence is that the field strengths are also constant provided one assumes in addition that $A^I_a$ are constant,
\be
F_{ab}= e^\mu_a e^\nu_b (\d_\mu e^c_\nu-\d_\nu e^c_\mu)A_c= c_{ab}^{\ \ c}A_c
\ee
Thus constant $c_{ab}^{\ \ c}$ gives regular geometries and is also necessary for getting the attractor equations.

\section{Gauged supergravity with one vector multiplet:}\label{appendix1}
In this section we will describe the supergravity model of \cite{Gunaydin:1999zx} used in $\S$\ref{constantanholonomy} in some detail. The field content of the $SO(1,1) \times \frac{SO(2,1)}{SO(2)}$ theory is:
\be 
\{e_\mu^a, \psi_\mu^i, A_\mu^I,B_{\mu\nu}^M, \lambda^{i\at}, \phi^{\xt} \}
\ee
where $i=1,2$ ; $\mu=0,\ldots,4$ ; $a=0,\ldots,4$ ; $I=0,1$ ; $M=2,3$ ; $\xt=1,2,3$ and $\at=0,1,2,3$. 

The scalar fields $\phi^{\xt}$ parametrise a very special manifold $\SS=SO(1,1) \times \frac{SO(2,1)}{SO(2)}$. The constraint \eqref{scalarcondition} written in a basis given by
\be
\xi^0= \frac{1}{\sqrt{2} ||\phi||^2} ;\ \ \xi^1=\phi^1 ;\ \ \xi^2=\phi^2 ;\ \xi^3=\phi^3 ,
\ee
takes the form
\be
N(\xi)=\sqrt{2} \xi^0 [ (\xi^1)^2-(\xi^2)^2-(\xi^3)^2 ]=1,
\ee
where,
\be
||\phi||^2=(\phi^1)^2-(\phi^2)^2-(\phi^3)^2
\ee
is assumed to be positive so that $a_{\It \Jt}$ and $g_{\xt \yt}$ are positive definite.  The $h^{\It}$ in \eqref{scalarcondition} is related to the above basis by $h^{\It}=\sqrt{\frac{2}{3}}\xi^{\It}|_{N=1}$ and the non-vanishing $C_{\It \Jt \Kt}$ are  $C_{011}=\frac{\sqrt{3}}{2}, C_{022}=C_{033}=-\frac{\sqrt{3}}{2}$.

For the computations, one also needs the vector/tensor metric $a_{\It \Jt}$ and the metric on the scalar manifold $g_{\xt \yt}$. We summarise them below:
\be
a_{\It \Jt}= 
\begin{pmatrix}
 ||\phi||^4 & 0 & 0 & 0 \\
0 &  2 (\phi^1)^2 ||\phi||^{-4}-||\phi||^{-2} & -2 \phi^1\phi^2 ||\phi||^{-4} & -2 \phi^1 \phi^3 ||\phi||^{-4}\\
0 & -2 \phi^1\phi^2 ||\phi||^{-4} & 2 (\phi^2)^2||\phi||^{-4}+||\phi||^{-2} & 2\phi^2\phi^3 ||\phi||^{-4}\\
0 & -2 \phi^1\phi^3 ||\phi||^{-4} & 2\phi^2\phi^3 ||\phi||^{-4} & 2 (\phi^3)^2 ||\phi||^{-4}+||\phi||^{-2}
\end{pmatrix}
\ee
\be
g_{\xt \yt}= \begin{pmatrix}
 4 (\phi^1)^2 ||\phi||^{-4}-||\phi||^{-2} & -4 \phi^1\phi^2 ||\phi||^{-4} & -4 \phi^1 \phi^3 ||\phi||^{-4}\\
-4 \phi^1\phi^2 ||\phi||^{-4} & 4 (\phi^2)^2||\phi||^{-4}+||\phi||^{-2} & 4\phi^2\phi^3 ||\phi||^{-4}\\
-4 \phi^1\phi^3 ||\phi||^{-4} & 4\phi^2\phi^3 ||\phi||^{-4} & 4 (\phi^3)^2 ||\phi||^{-4}+||\phi||^{-2}
\end{pmatrix}
\ee
\section{Field equations:}\label{bianchi26fieldeq}
In this section we summarise the field equations for the Bianchi Type II and VI metrics.
\subsection{Bianchi type II:}
The diagonal components of the Einstein equations $(tt,rr,xx,yy,zz)$ are:
\begin{align}
\begin{split}
 (1+12 v^2 +20 vw +12 w^2)(\phi^1_c)^4 + 2 (A^{0\bar{0}})^2 (3 g^2 L^2+u^2(\phi^1_c)^8)-48 L^2 g_R^2 V_0^2 &=0 
\end{split}\nn\\ 
\begin{split}
-(1+ 4 v^2 +12 vw +4w^2+8uv+8uw)(\phi^1_c)^4 + (A^{0\bar{0}})^2 (6 g^2 L^2- 2 u^2(\phi^1_c)^8)\\ + 48 L^2 g_R^2 V_0^2 &=0 
\end{split}\nn\\
\begin{split}
-(-1+ 12v^2+12vw+4w^2+8uv+4uw+4u^2)(\phi^1_c)^4+ 2(A^{0\bar{0}})^2 (3 g^2 L^2+u^2(\phi^1_c)^8)\\+ 48 L^2 g_R^2 V_0^2 &=0
\end{split}\nn \\
\begin{split}
-(3+4v^2+4w^2+4uw+4vw+4uv+4u^2)(\phi^1_c)^4+ 2(A^{0\bar{0}})^2 (3 g^2 L^2+u^2(\phi^1_c)^8)\\+48 L^2 g_R^2 V_0^2 &=0
\end{split}\nn\\
\begin{split}
-(-1+4v^2+12w^2+8uw+4u^2+12vw+4uv)+ 2(A^{0\bar{0}})^2 (3 g^2 L^2+u^2(\phi^1_c)^8)\\+48 L^2 g_R^2 V_0^2 &=0
\end{split}
\end{align}
The gauge field equation for $A^{0\bar{0}}$ is 
\be
\hat{e} A^{0\bar{0}}(3 g^2 L^2 - 2 u(v+w) (\phi^1_c)^8=0.
\ee
The scalar field equation is
\be
3 g^2 L^2 - u^2 (\phi^1_c)^8=0.
\ee
The $xx$ and $zz$ components of the Einstein equations together with the gauge field and the scalar field equations give,
\be
u=\sqrt{3} \frac{g L}{(\phi^1_c)^4};\quad v=w=\frac{u}{4}.
\ee
Substituting this into the Einstein equations, the $(tt,rr,xx,yy)$\footnote{The $zz$ equation is same as the $xx$ equation. } equations are given by
\begin{align}
12 (A^{0\bar{0}})^2 g^2 L^2 + \frac{33 g^2 L^2}{4(\phi^1_c)^4}+(\phi^1_c)^4-48 L^2 g_R^2 V_0^2&=0\nn\\
\frac{-63 g^2 L^2}{4(\phi^1_c)^4}-(\phi^1_c)^4+48 L^2 g_R^2 V_0^2&=0\nn\\
12 (A^{0\bar{0}})^2 g^2 L^2 - \frac{105 g^2 L^2}{4(\phi^1_c)^4}+(\phi^1_c)^4+48 L^2 g_R^2 V_0^2&=0\nn\\
12 (A^{0\bar{0}})^2 g^2 L^2 - \frac{81 g^2 L^2}{4(\phi^1_c)^4}-3(\phi^1_c)^4+48 L^2 g_R^2 V_0^2&=0
\end{align}
which can be solved to get
\be
A^{0\bar{0}}=\sqrt{\frac{5}{8}}\frac{1}{(\phi^1_c)^2} ; \quad L= \sqrt{\frac{2}{3}}\frac{(\phi^1_c)^4}{g},
\ee
with the constraint,
\be
\frac{23}{2(\phi^1_c)^4}=32 \frac{g_R^2}{g^2}V_0^2.
\ee
\subsection{Bianchi Type VI}
For this case, the off-diagonal $rx$ Einstein equation gives the condition,
\be
v=-w h
\ee
The rest of the Einstein equations $(tt,rr,xx,yy,zz)$ are,
\begin{align}
\begin{split}
2(1+h+h^2+w^2(1-h+h^2))(\phi^1_c)^4+ (A^{0\bar{0}})^2(3 g^2 L^2 + u^2 (\phi^1_c)^8)-24 L^2 g_R^2 V_0^2 &=0
\end{split}\nn\\
\begin{split}
 -2(1+uw+h-w(w+u)h+h^2)(\phi^1_c)^4+ (A^{0\bar{0}})^2(3 g^2 L^2 - u^2 (\phi^1_c)^8)+24 L^2 g_R^2 V_0^2 &=0
\end{split}\nn\\
\begin{split}
 -2(w^2+uw+u^2+h-w(w+u)h+w^2h^2)(\phi^1_c)^4+(A^{0\bar{0}})^2(3 g^2 L^2 + u^2 (\phi^1_c)^8)\\+24 L^2 g_R^2 V_0^2 &=0
\end{split}\nn\\
\begin{split}
 -2(w^2+uw+u^2+h^2)(\phi^1_c)^4+(A^{0\bar{0}})^2(3 g^2 L^2 + u^2 (\phi^1_c)^8)+24 L^2 g_R^2 V_0^2 &=0
\end{split}\nn\\
\begin{split}
 -2(1+u^2-uwh+w^2h^2)(\phi^1_c)^4+(A^{0\bar{0}})^2(3 g^2 L^2 + u^2 (\phi^1_c)^8)+24 L^2 g_R^2 V_0^2 &=0
\end{split}
\end{align}
The gauge field equation for $A^{0\bar{0}}$ is given by,
\be
3 g^2 L^2 +u w(-1+h)(\phi^1_c)^8=0.
\ee
The scalar field equation reduces to,
\be
3 g^2 L^2- u^2 (\phi^1_c)^8=0
\ee
These two equations can be solved (assuming $h\neq 1$) to get,
\be
u= \sqrt{3} \frac{g L}{(\phi^1_c)^4};\quad w=\frac{u}{1-h}. 
\ee
The remaining Einstein equations are all not independent and can be solved to get,
\be
L= \frac{(\phi^1_c)^4}{\sqrt{6}g} |h-1|; \quad A^{0\bar{0}}=\sqrt{\frac{-2h}{(-1+h)^2}}\frac{1}{(\phi^1_c)^2}.
\ee
The gauge field solution implies that $h<0$ for $A^{0\bar{0}}$ to be real. The constraint on the free parameters is given by,
\be
 \frac{8(3-h+h^2)}{(\phi^1_c)^4(h-1)^2}= 32 \frac{g_R^2}{g^2}V_0^2.
\ee

\providecommand{\href}[2]{#2}\begingroup\raggedright\endgroup

\end{document}